\documentclass[twocolumn,10pt]{IEEEtran}%
\usepackage[utf8]{inputenc}
\usepackage{amsfonts}
\usepackage{amsmath}
\usepackage{multicol}
\usepackage{amssymb}
\usepackage{graphicx}
\usepackage{mathtools}
\usepackage{xcolor}
\usepackage{ntheorem}
\usepackage{algorithm}
\usepackage{algorithmicx}
\usepackage{algpseudocode}
\usepackage{setspace}
\usepackage{listings}
\usepackage{cite}
\usepackage{url}
\usepackage{enumitem}

\algnewcommand\algorithmicoutput{\textbf{Output:}} 
\algnewcommand\Output{\item[\algorithmicoutput]}
\algnewcommand\algorithmicinput{\textbf{Input:}} 
\algnewcommand\Input{\item[\algorithmicinput]}

\def\baselinestretch{1}

\begin{document}
\title{An Intuitive Derivation of the Coherence Index \\ Relation in Compressive Sensing}

\author{
Ljubi\v{s}a~Stankovi\'{c},~\IEEEmembership{Fellow,~IEEE,}
Danilo Mandic,~\IEEEmembership{Fellow,~IEEE,} 
Milo\v{s} Dakovi\'{c},~\IEEEmembership{Member,~IEEE},
Ilya Kisil
\thanks{
L. Stankovi\'{c} and M. Dakovi\'{c} are with the  University of
Montenegro, Podgorica, Montenegro. D. Mandic and I. Kisil are with the Imperial College London, London, United Kingdom.
Contact e-mail: ljubisa@ac.me}
}
\maketitle

\setcounter{tocdepth}{3}

\begin{abstract}
The existence and uniqueness conditions are a prerequisite for reliable reconstruction of sparse signals from reduced sets of measurements within the Compressive Sensing (CS) paradigm. However, despite their underpinning role for practical applications, existing uniqueness relations are either computationally prohibitive to implement (Restricted Isometry Property), or involve mathematical tools that are beyond the standard background of engineering graduates (Coherence Index). This can introduce conceptual and computational obstacles in the development of engineering intuition, design of suboptimal practical solutions, or understanding of limitations. To this end, we introduce a simple but rigorous derivation of the coherence index condition, based on standard linear algebra, with the aim to empower signal processing practitioners with intuition in the design and ease in implementation of CS systems. Given that the coherence index is one of very few CS metrics that admits mathematically tractable and computationally feasible calculation, it is our hope that this work will help bridge the gap between the theory and applications of compressive sensing.  
\end{abstract}

\section{Introduction and Basic CS Setting}

Compressive Sensing (CS) is a maturing field which, under appropriate conditions, provides a rigorous framework for efficient data acquisition. Examples include the recovery of sparse signals from vastly reduced sets of measurements, applications which rest upon reliable sensing from the lowest possible number of measurements, and practical solutions when some measurements are physically unavailable or heavily corrupted by disturbance. 

Research under an overarching umbrella of sparsity has been a major topic of investigation for about a quarter of century, and has produced solid theory  to support the exact reconstruction of  sparse signals in compressive sensing scenarios. Among the uniqueness tools in CS, the coherence index is of particular interest for practitioners as it is one of the few supporting theoretical  tools in compressive sensing which can be calculated in a computationally feasible way, and is the focus of this work. However, its derivation follows a rather complex and convolved path which is beyond the standard background of an engineering graduate; this spurred us to revisit the coherence index from a signal processing perspective, in order to equip practitioners with intuition in the design and ease  of interpretation in the analysis. We also provide an intuition behind  the feasibility of the calculation of the Restricted Isometry Property (RIP) condition.

\subsection{Definitions and Notation}

\textit{Definition 1:} A sequence  $\{X(k)\}$, $k=0,1,\dots,N-1$ is referred to as a sparse sequence, if the number $K$ of its nonzero elements, $X(k)\ne 0$, is much smaller than its total length, $N$, that is,
$$X(k) \ne 0 \textrm{ for } k \in \{k_1,k_2,\dots,k_K\}, \,\, K \ll N.$$

\textit{Defintion 2:} A linear combination of elements of $X(k)$, given by
\begin{gather}
y(m)=\sum_{k=0}^{N-1} a_m(k)X(k), \label{MeasDef}
\end{gather}  
is called a measurement, with the weighting coefficients (weights)  denoted by $a_m(k)$.  

\begin{figure}[tbh]
	\centering
	\includegraphics[]{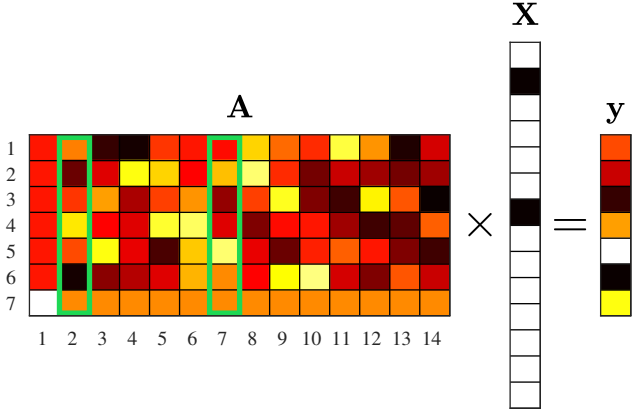}
	\caption{Principle of compressive sensing. The short and wide measurement matrix $\mathbf{A}$ maps the original $N$-dimensional  $K$-sparse vector, $\mathbf{X}$, to an $M$-dimensional dense vector of measurements, $\mathbf{y}$, with $M<N$ and $K \ll N$. In our case $N=14$, $M=7$, and $K=2$.  Since $M$ is the maximum rank of $\mathbf{A}$ and $M<N$, the original $N$-dimensional vector $\mathbf{X}$ cannot, in general, be recovered from the measurements, $\mathbf{y}$  (lack of degrees of freedom). However, the CS paradigm allows for the complete and unique recovery of sparse signals, with the coherence index being a common way to define the corresponding unique recovery conditions. The same $\mathbf{A}$ is used in Fig. \ref{slika_PPP_1}.}
	\label{PrincIlli}
\end{figure}

The above sensing scheme produces measurements, $y(m)$, $m=0,1,\dots,M-1$, and admits a vector/matrix form given by 
\begin{gather}
\mathbf{y}=\mathbf{A}\mathbf{X},  \label{mmmess}
\end{gather}  
where $\mathbf{y}=\{y(m)\}$ is an $M\times 1$ column vector of the measurements, $\mathbf{A}$ is an $M\times N$ measurement matrix which comprises the weights $a_m(k)$ as its elements, and $\mathbf{X}$ is an $N\times 1$ sparse column vector with elements $X(k)$. An illustration of the CS concept is given in Fig. \ref{PrincIlli}.

Without loss of generality, we shall assume that the measurement matrix, $ \mathbf{A}$, is normalized, so that the energy of its columns sums up to unity. Consequently, the diagonal elements of its symmetric Gram form, $\mathbf{A}^H \mathbf{A}$, are equal to $1$, where $\mathbf{A}^H$ is the complex conjugate transpose of $ \mathbf{A}$.

While compressive sensing theory states that, under certain mild conditions, it is possible to reconstruct a sparse $N$-dimensional vector, $\mathbf{X}$, from a reduced $M$-dimensional set of measurements, $\mathbf{y}$, for the applications of CS to become more widespread, the practitioners require physically meaningful, intuitive, and easily interpretable uniqueness tools - a subject of this work.

\section{A Solution to the CS Paradigm}
 Several approaches have been established for the CS paradigm, and we here  follow the principles behind the matching pursuit approach. The simplest case, when the positions of nonzero elements in $\mathbf{X}$  are known, is considered first to provide both the intuition and an example for the detection of unknown positions of the nonzero elements in $\mathbf{X}$. This  then serves as a basis for our derivation of the uniqueness relation through coherence index,  for a general case of unknown positions of the nonzero elements in $\mathbf{X}$.         
\subsection{Known Coefficient Positions}
Consider the case with $K$ nonzero elements of $\mathbf{X}$ located at arbitrary but known positions, that is, $X(k) \ne 0$  for $k \in \{k_1,k_2,\dots,k_K\}$.
Compared to the general form of the measurement relations in (\ref{mmmess}), this gives rise to the following reduced system of equations
\begin{equation}
\!\!\left[
\begin{matrix} 
y(0)\\
y(1)\\
\vdots\\
y(M-1)\\
\end{matrix}
\right]
\!\!\!=\!\!\!
\left[
\begin{matrix} 
a_{0}(k_1) &  \dots & a_{0}(k_{K}) \\
a_{1}(k_1) &  \dots & a_{1}(k_{K})\\
\vdots & \ddots  & \vdots \\
a_{M-1}(k_1) & \dots & a_{M-1}(k_{K})
\end{matrix} 
\right] \!\! \!\!
\left[
\begin{matrix} 
X(k_1)\\
X(k_2)\\
\vdots\\
X(k_K)\\
\end{matrix}
\right]. \!\label{RndKnownposi}
\end{equation}  
For a succeful CS recovery, the matrix form of the above system, given by 
$$\mathbf{y}=\mathbf{A}_{MK}\mathbf{X}_K,$$
 need to be solved for the nonzero elements $X(k)$, located at $k \in \{k_1,k_2,\dots,k_K\}$, which are here conveniently grouped into a $K \times 1$ vector $\mathbf{X}_K$. 
Observe that the matrix $\mathbf{A}_{MK}$ is an $M \times K$ dimensional sub-matrix of the full $M \times N$ measurement matrix $\mathbf{A}$ in (\ref{mmmess}), whereby only the columns that correspond to the positions of the nonzero elements in $\mathbf{X}$ are kept (for illustration, see column 2 and column 7 in Fig. \ref{PrincIlli}). 
The smallest number of measurements needed to recover the $K$-element vector  $\mathbf{X}_K$ is therefore $M=K<N$. For $M>K$, as in Fig. \ref{PrincIlli}, the system is overdetermined and the solution is found in the Least Squares  (LS) sense, to give 
\begin{equation}\mathbf{X}_K=(\mathbf{A}^H_{MK}\mathbf{A}_{MK})^{-1}\mathbf{A}_{MK}^H\mathbf{y}=\textrm{pinv}(\mathbf{A}_{MK})\mathbf{y}, \label{pinvSol}
\end{equation} 
where $\textrm{pinv}(\mathbf{A}_{MK})=(\mathbf{A}^H_{MK}\mathbf{A}_{MK})^{-1}\mathbf{A}_{MK}^H$ denotes the pseudo-inverse of the matrix $\mathbf{A}_{MK}$,  while the matrix $\mathbf{A}^H_{MK}\mathbf{A}_{MK}$  is referred to as the $K \times K$ dimensional Gram matrix of $\mathbf{A}_{MK}$.

 From (\ref{pinvSol}),  the existence of a recovery solution requires that the inverse $(\mathbf{A}^H_{MK}\mathbf{A}_{MK})^{-1}$ does exist, or in other words that
 $ \textrm{rank}(\mathbf{A}^H_{MK}\mathbf{A}_{MK})=K$.
  This requirement can be equally expressed via the matrix condition number, as    
 $\textrm{cond}(\mathbf{A}^H_{MK}\mathbf{A}_{MK})<\infty$, which casts the existence condition into a more convenient form dictated by the eigenvalue spread of $\mathbf{A}^H_{MK}\mathbf{A}_{MK}$.  
 For noisy measurements, the reconstruction error comprises contributions from both input noise and the ill-posedness due to a high matrix condition number (``mathematical artefact''). In this sense, the goal of a successful  compressive sensing (sampling) strategy can be interpreted as that of forming the measurement matrix, $\mathbf{A}$, in such a way to produce the condition number as close to $1$ as possible.

\subsection{Signal Processing (DFT) Framework for CS} 
The most fertile domains to account for signal sparsity are common linear signal transforms, as these allow for original time-domain samples of the signal $x(n)$ to be considered as measurements (linear combinations) of the representation domain coefficients, $X(k)$. For example, when the Discrete Fourier Transform (DFT) is used as the signal sparsity domain, the measured signal samples can be expressed as   
\begin{gather}
y(m)=x(n_m)=\frac{1}{\sqrt{M}}\sum_{k=0}^{N-1}  e^{j 2\pi n_mk /N}X(k), \label{DFTpart}
\end{gather}
for  $m=0,1,\dots,M-1$. In this case, the measurements, $y(m)$, can be regarded as a \textit{reduced set of signal samples}, $x(n_m)$.  Observe that this expression is conformal with the general CS formulation  (\ref{MeasDef}), where the weights $a_m(k)=\exp(j 2\pi n_mk /N)/\sqrt{M}$. Since CS employs a random subset of time instants $\{n_1,n_2,\dots,n_M\} \subset \{0,1,2,\dots,N-1\}$, the measurement matrix $\mathbf{A}$ is obtained from the inverse DFT transform matrix, whereby only the rows corresponding to $\{n_1,n_2,\dots,n_M\}$ are kept. The matrix $\mathbf{A}$, obtained in such a way, is called a partial DFT measurement matrix.     

\textit{Remark 1:} The convenience of the considered DFT representation allows us to consider the Nyquist sub-sampling as a special case of a reduced set of measurements, with the specific positions of $K$ nonzero elements given by 
$$\mathbf{X}=[X(0),X(1),\dots,X(K-1),0,0,\dots,0]^T.$$ 
The classic sub-sampling operation can then be explained starting from the assumption that only the first $K=N/P$ elements are nonzero, with $P>1$  an integer.  Then, the original signal can be sub-sampled at $n_m=mP$, since this yields  $a_m(k)=\exp(j 2\pi mk /K)/\sqrt{K}$ with $\mathbf{A}^H_{MK}\mathbf{A}_{MK}$ being an identity matrix.

Some important applications, such as radar signal processing, routinely deal with a very small number of nonzero elements, $X(k)$. However, in contrast to the classic sub-sampling scenario these nonzero transform-domain elements may be located at any position $k \in\{0,1,\dots,N-1\}$. While this makes it impossible to perform classical sub-sampling, this class of applications admits a unique solution with a reduced number of signal samples, within the CS theory framework. 

Despite obvious methodological advantages of CS over classical analyses, operation on a reduced set of measurements compromises the uniqueness of the CS solution. It is therefore natural to first examine which conditions should be satisfied by the measurements (both their number and properties of the measurement matrix) so that the existence of the CS solution is guaranteed and the solution is unique. 

\textit{Example 1:} In order to illustrate how a reduced set of samples can compromise uniqueness of the solution, consider a signal with the total length of $N$ and the simplest sparsity degree of $K=1$ in the DFT domain, that is, with only one nonzero element in $\mathbf{X}$.  Assume that a significant number of $M=N/2$ measurements (signal samples) are available at  $n=0,2,\dots,N-4,N-2$, with an even $N$, and assume that the measurement values are $y(m)=x(2m)=1$. The solution to this simple problem is then not unique, since $x(n)=\exp(j2\pi n k_1 /N)$, $n=0,1,\dots,N-1$, for both $k_1=0$ and $k_1=N/2$, satisfies all problem conditions. 
 
 \noindent\textbf{Uniqueness of the CS paradigm:} In general, the question of uniqueness can be considered within the following framework.  Consider a $K$-sparse vector  $\mathbf{X}$, with the nonzero elements $X(k)\ne 0$ at $k \in \{k_1,k_2,\dots,k_K\}$, and assume that its vector form, $\mathbf{X}_K$, is a solution to $\mathbf{y}=\mathbf{A}_{MK}\mathbf{X}_K.$ Assume also that the solution is not unique, so that there exists another vector, $\mathbf{X}'$, with nonzero elements at different positions $k \in \{k_{K+1},k_{K+2},\dots,k_{2K}\}$, whose reduced form  $\mathbf{X}'_K$, supports the same measurements $\mathbf{y}$, that is,  $\mathbf{y}=\mathbf{A}'_{MK}\mathbf{X}'_K$.  Then, $\mathbf{A}_{MK}\mathbf{X}_K-\mathbf{A}'_{MK}\mathbf{X}'_K=\mathbf{0}$, and this matrix equation can be combined  into $\mathbf{A}_{M2K}\mathbf{X}_{2K}=\mathbf{0}$, where $\mathbf{A}_{M2K}$ is an $M\times 2K$-dimensional sub-matrix of the measurement matrix $\mathbf{A}$ and $\mathbf{X}_{2K}$ is a $2K$-dimensional vector. Then,
 \begin{itemize}
  \item 
  A nontrivial solution of the matrix equation $\mathbf{A}_{M2K}\mathbf{X}_{2K}=\mathbf{0}$ indicates that the CS solution is nonunique. The condition for a nonunique solution is therefore $ \textrm{rank}(\mathbf{A}^H_{M2K}\mathbf{A}_{M2K})<2K$, for at least one combination of $2K$ nonzero element positions. 
 
 \item 
  If $ \textrm{rank}(\mathbf{A}^H_{M2K}\mathbf{A}_{M2K})=2K$, for all possible combinations of $2K$ nonzero element positions (out of $N$), the scenario of two $K$-sparse solutions is not possible and the solution is unique. 
 \end{itemize}
 	
 The above rationale  is a starting point for the definition of common uniqueness criteria in CS. The coherence index relation (whose simple derivation and interpretation is the subject of this paper) is typically derived through the Gershogorin disk theorem \cite{GSH}.  
 
 The maximum robustness of the  condition $ \textrm{rank}(\mathbf{A}^H_{M2K}\mathbf{A}_{M2K})=2K$ is achieved if $\textrm{cond}(\mathbf{A}^H_{MK}\mathbf{A}_{MK})$ is close to unity. It should be mentioned that the RIP is tested in a similar way, by a combinatorial consideration of the matrix norm 
 $\left\| \mathbf{A}_{MK} \mathbf{X}_{2K}\right\|_2^2/\left\|\mathbf{X}_{2K}\right\|_2^2 $ through  eigenvalue analysis. 
 
The general case, with unknown positions of the nonzero elements in $\mathbf{X}$, can be solved by combining the above described method for the known positions and a direct search approach, as described in Sidebar 1. However, this is not computationally feasible.
 
\begin{figure*}[thb]
	\centering
	\fboxsep=1pt
	\fbox{\includegraphics[]{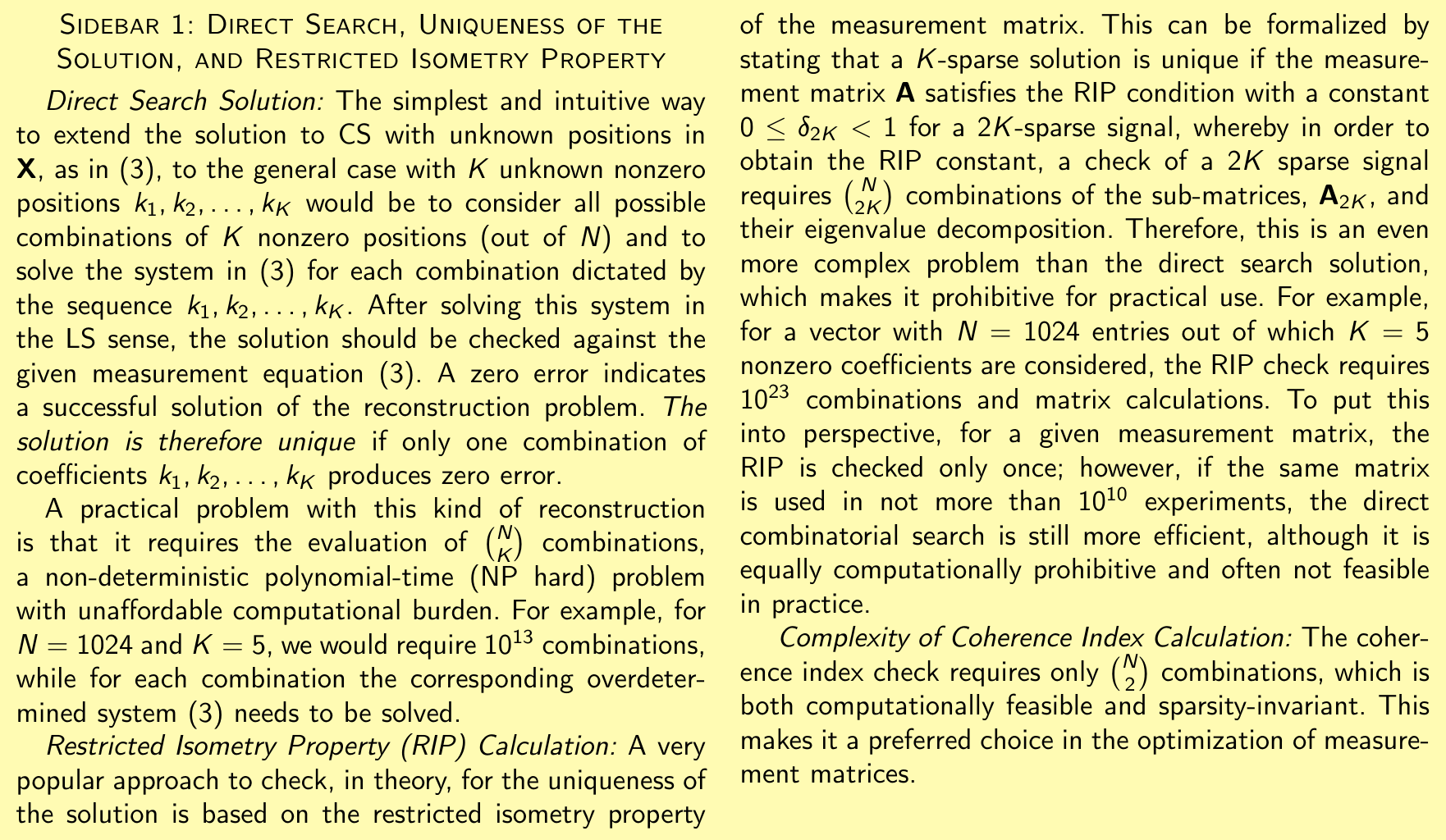}}
\end{figure*}

\subsection{Detection of Unknown Coefficient Positions}

In the CS setup, the positions of nonzero elements in $\mathbf{X}$ are typically not known. Then, a  natural approach to solve the CS reconstruction problem would be to adopt a two-step strategy as follows: 
\begin{itemize}
	\item[] Step 1: Detect the positions of nonzero elements,
	\item[] Step 2: Apply an algorithm for reconstruction with known nonzero element positions. 
\end{itemize}	

An intuition for the estimate of the positions of nonzero elements in Step 1
comes from the \textit{linear} nature of measurements, $y(m)$,  obtained as linear combinations of the sparsity domain elements, $X(k)$, and the corresponding rows of the measurement matrix, $\mathbf{A}$. 

\textit{Remark 2:} The linearity of the CS paradigm in (\ref{mmmess}), admits a back-projection of the measurements, $\mathbf{y}$, to the measurement matrix, $\mathbf{A}$, defined by
\begin{equation}
\mathbf{X}_0=\mathbf{A}^H \mathbf{y}=\mathbf{A}^H \mathbf{A} \mathbf{X}\label{Initi}
\end{equation}
  to be used to estimate the positions of nonzero elements in $\mathbf{X}$.  
In an ideal case, the matrix $\mathbf{A}^H \mathbf{A}$ should ensure that the initial estimate, $\mathbf{X}_0$, contains exactly $K$ elements at postions $\{k_1,k_2,\dots,k_K\}$ for which the magnitudes are larger than the biggest magnitude at the remaining positions. Then, by taking the positions of these highest magnitude elements in $\mathbf{X}_0$ as the set $\{k_1,k_2,\dots,k_K\}$ in (\ref{RndKnownposi}), the algorithm for the known nonzero element positions, from the previous section, can be applied to reconstruct the signal. 

\textit{Remark 3:} Note that if $\mathbf{A}^H \mathbf{A}$ were an identity matrix, the initial estimate, $\mathbf{X}_0$, would correspond to the exact solution, $\mathbf{X}$. However, with
a reduced number of measurements, $M<N$, this cannot be achieved (due to the Welch lower bound). A pragmatic requirement for the existence of the CS solution would therefore be
 that the off-diagonal elements of $\mathbf{A}^H \mathbf{A}$ are as small as possible compared to its unit diagonal 
elements.  

The condition that all $K$ elements in the initial estimate, $\mathbf{X}_0$,  which are located at the nonzero element positions in the original sparse vector, $\mathbf{X}$,  are larger than any other element in the initial estimate $\mathbf{X}_0$ can be relaxed through an iterative procedure. To be able to find the position $k_1$ of the largest nonzero element in $\mathbf{X}_0$, its value must be larger than the values $X_0(k)$ at the original zero-valued element positions. After the largest element position is found and its value is estimated this component can be reconstructed and subtracted from measurements, $\mathbf{y}$, and the procedure is continued with the remaining $(K-1)$-sparse elements, in an iterative manner. The stopping criterion then becomes that $\mathbf{A}_{MK}\mathbf{X}_K=\mathbf{y}$ should hold for the estimated nonzero positions $\{k_1,k_2,\dots,k_K\}$ and elements $X(k)$, as outlined in Algorithm 1.    

\begin{algorithm}[!tbh]
	\caption{\!\!\textbf{.} \, Matching Pursuit Based Reconstruction}
	\label{Norm0Alg}
	\begin{algorithmic}[1]
		\Input
		\Statex
		\begin{itemize}
			\item Measurement vector $\mathbf{y}$
			\item Measurement matrix $\mathbf{A}$
			\item Required precision $\varepsilon$
		\end{itemize}
		\Statex
		\State $\mathbb{K} \gets \emptyset$
		\State $\mathbf{e} \gets \mathbf{y} $
		\While{$\left\| \mathbf{e} \right\|_2 > \varepsilon$}
		\State $k \gets $ position of 
			the highest 
			value in $\mathbf{A}^H\mathbf{e}$
		\smallskip
		\State $\mathbb{K} \gets \mathbb{K} \cup k $
		\State $\mathbf{A}_K \gets $ columns of matrix $\mathbf{A}$ selected by 
		set $\mathbb{K} $
		\State $\mathbf{X}_K \gets \operatorname{pinv}(\mathbf{A}_K)\mathbf{y}$
		\State $\mathbf{y}_K \gets \mathbf{A}_K\mathbf{X}_K$
		\State $\mathbf{e} \gets \mathbf{y} -  \mathbf{y}_K$
		\EndWhile
		\smallskip
		\State $\displaystyle \mathbf{X} \gets
		\begin{cases} \mathbf{0}, & \text{for the positions not in }\mathbb{K}, \\
		\mathbf{X}_K, & \text{for the positions in }\mathbb{K}.
		\end{cases}$
		
		\Statex
		\Output
		\Statex
		\begin{itemize}
			\item Reconstructed signal  elements $\mathbf{X}$
		\end{itemize}
	\end{algorithmic}
\end{algorithm}

\section{Unique Reconstruction Condition}

The key criterion for signal reconstruction with a reduced set of measurements is the uniqueness of the result. In CS methodology, the uniqueness is commonly defined through the coherence index, as stated bellow.

\noindent\textbf{Proposition: } \textit{The reconstruction of a $K$-sparse signal, $\mathbf{X}$, is unique if the coherence index, $\mu$, of the measurement matrix,   $\mathbf{A}$, satisfies \cite{donoho2006}
\begin{equation}
K < \frac{1}{2}\left(1+\frac{1}{\mu}\right), \label{Kkoc}
\end{equation}
where for a normalized measurement matrix, $\mathbf{A}$, the coherence index is equal to the maximum absolute off-diagonal element of $\mathbf{A}^H \mathbf{A}$.}

The above proposition is usually proven based on the Gershogorin disk theorem \cite{GSH}, a topic not covered in engineering curricula; this is an obstacle which prevents both more widespread engagement of engineers in the CS field and success of eventual applications. To this end, 
we shall now derive the reconstruction condition in a self-contained and intuitive way which does not require advanced mathematics. 

\begin{figure*}[tbh]
	\centering
	\includegraphics[]{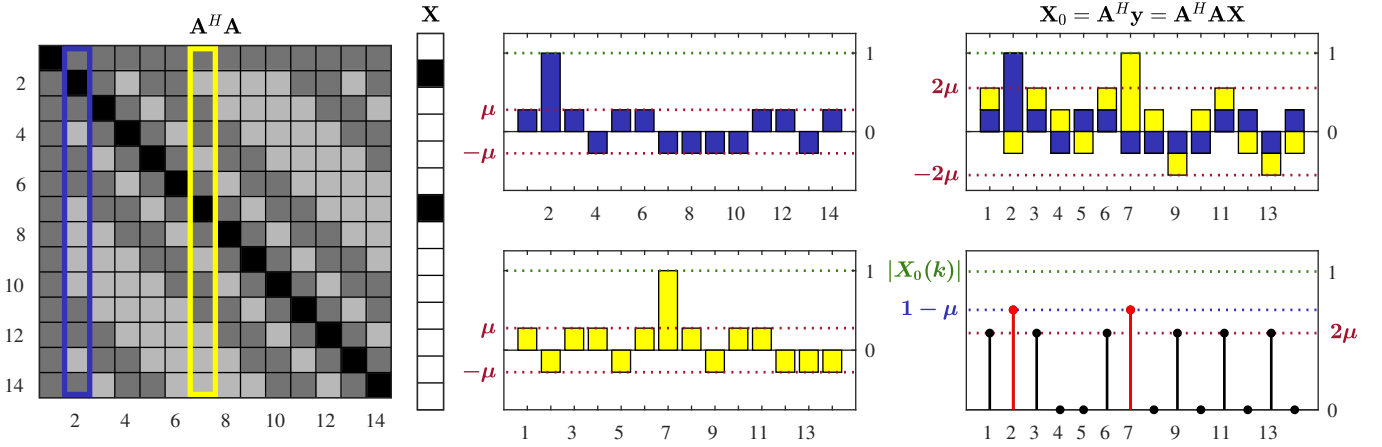}
	\caption{Illustration of the coherence index relation for $K=2$. Left: Matrix $\mathbf{A}^H\mathbf{A}$ for the equiangular tight frame (ETF) matrix $\mathbf{A}$ from Fig. \ref{PrincIlli}, and the original elements in $\mathbf{X}$ with nonzero values at $k_1=2$ and $k_2=7$. The values in matrix $\mathbf{A}^H\mathbf{A}$ are as follows: black=1, white=0, dark gray=$+\mu$, and light gray=$-\mu$. Middle: Components in the initial estimate $\mathbf{X}_0=\mathbf{A}^H\mathbf{y}$, calculated using measurements $\mathbf{y}=\mathbf{AX}$ resulting from the nonzero element at $k_1=2$ in the blue barplot and from the nonzero element at $k_2=7$ in the yellow barplot. Right: Visual demonstration of the condition for a correct detection of the original nonzero element position from the initial estimate is $1-\mu>2\mu$, with the stacked barplot for both components in the top panel and the resulting $|X_0(k)|$ in the bottom panel. }
	\label{slika_PPP_1}
\end{figure*}

As an example, consider a $7 \times 14$ measurement matrix $\mathbf{A}$; for clarity we employ the so called equiangular tight frame (ETF) matrix, for which the absolute value of the off-diagonal elements of $\mathbf{A}^H \mathbf{A}$ is constant and equal to the coherence index, $\mu$. The matrix $\mathbf{A}^H \mathbf{A}$ is visualized in Fig. \ref{slika_PPP_1} (left). Its diagonal elements are, by definition, equal to $1$, while its off-diagonal elements that have values $\pm \mu$ can be treated as the disturbances in the detection of nonzero element positions in $\mathbf{X}$. Observe that the off-diagonal column elements of this matrix represent the normalized contribution of the corresponding nonzero element in the sparse vector $\mathbf{X}$ to the cumulative disturbance.

\begin{figure}[phtb]
	\centering
	\includegraphics[scale=1]{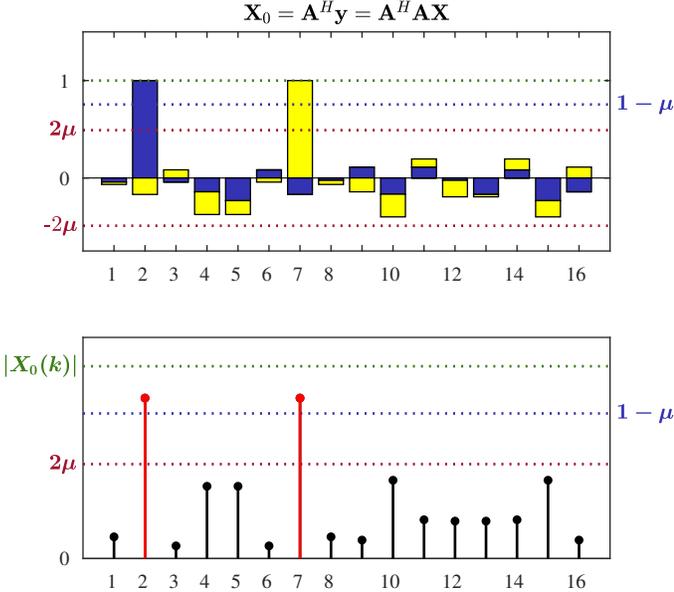}
	
	\caption{Initial estimate, $\mathbf{X}_0$, for a $12 \times 16$ partial DFT measurement matrix,  with $\mu=0.2455$ and the sparsity degree of $K=2$ in $\mathbf{X}$, with $X(2)=X(7)=1$ (stacked barplot in the top panel). The positions of nonzero elements in $\mathbf{X}$ are detected at $k \in \{2,7\}$ if their amplitude, reduced by the maximum possible disturbance from the other component, $\mu$, is above the maximum possible disturbance, $2\mu$, of both components, at a position of an original zero-valued element in $\mathbf{X}$, $k \notin \{2,7\}$ (bottom panel). Observe a larger misdetection margin compared to the ETF case, indicated by the corresponding $1-\mu$ and $2\mu$ levels, which is a consequence of different distributions of disturbances (off-diagonal elements of matrix $\mathbf{A}^H\mathbf{A}$).}
	\label{coh_text4_1}
\end{figure}

\begin{figure*}[tbhp]
	\centering
	
	
	\includegraphics[]{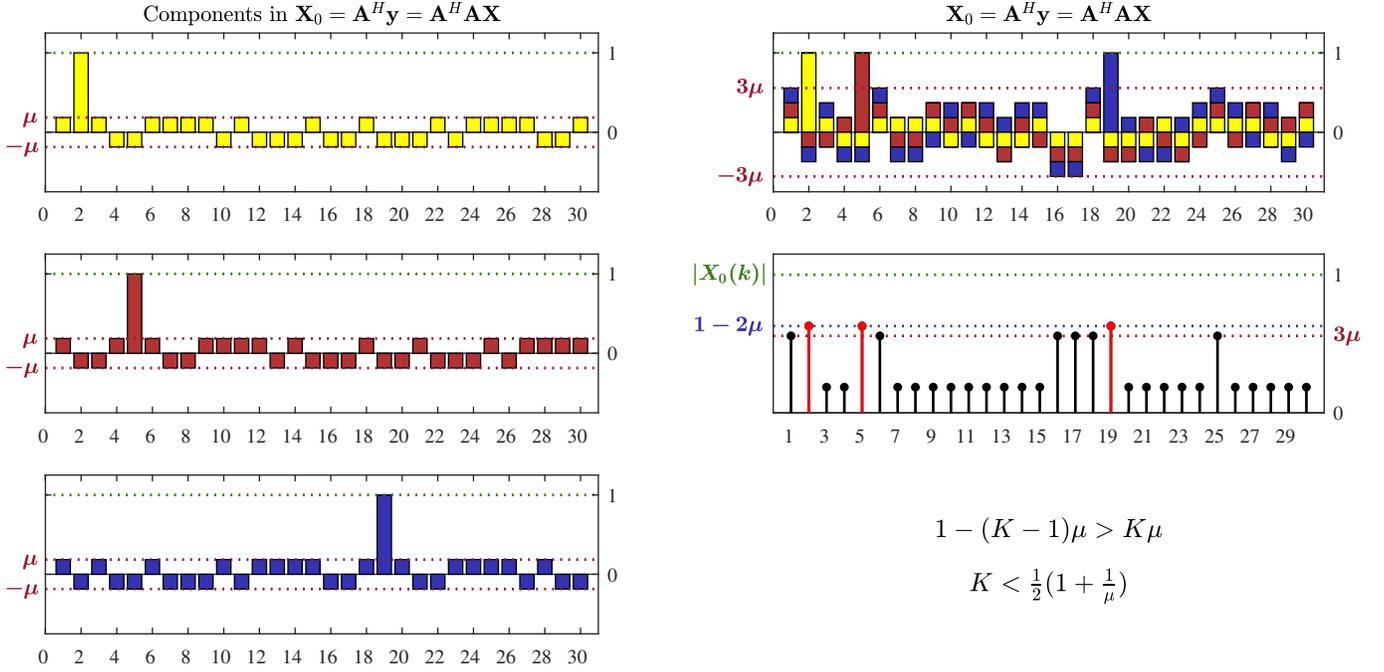}
	
	\caption{Illustration of  the coherence index relation for $K=3$. Left: Components of the initial estimate $\mathbf{X}_0=\mathbf{A}^H \mathbf{y}=\mathbf{A}^H \mathbf{A} \mathbf{X}$ resulting from each of three unit-valued nonzero elements in  $\mathbf{X}$ at  $k \in \{2,5,19\}$. Right: These components are combined into the initial estimate $\mathbf{X}_0$.  The positions of the nonzero elements in $\mathbf{X}$ are detected at $k \in \{2,5,19\}$, using the absolute value of the initial estimate, $|X_0(k)|$, if their amplitudes, reduced for a maximum possible disturbance, $2\mu$, from the other two components, are above the maximum possible disturbance, $3\mu$, of all three  components, at a position of an original zero-valued element in $\mathbf{X}$, $k \notin \{2,5,19\}$. The condition for the correct detection of the nonzero element positions, with $K=3$,  is therefore $1-2\mu>3\mu$, that is $1-(K-1)\mu>K\mu$, which straightforwardly generalizes to any $K$.}
	\label{slika_PPP_2}
\end{figure*}

\begin{figure*}[tbhp]
	\centering
	\fboxsep=1pt
	\fbox{\includegraphics[]{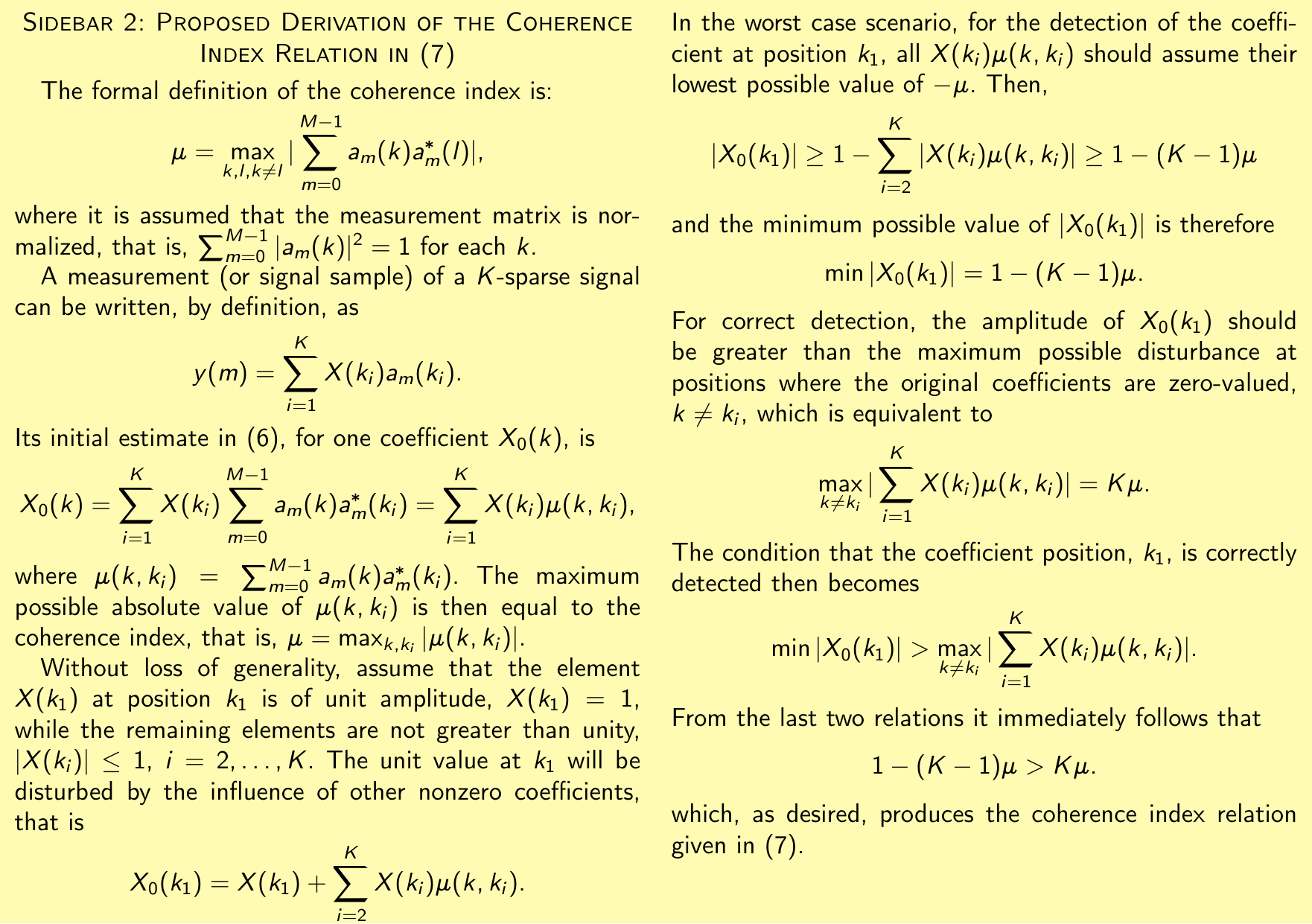}}
\end{figure*}

\begin{figure*}[tbp]
	\centering
	\fboxsep=1pt
	\fbox{\includegraphics[]{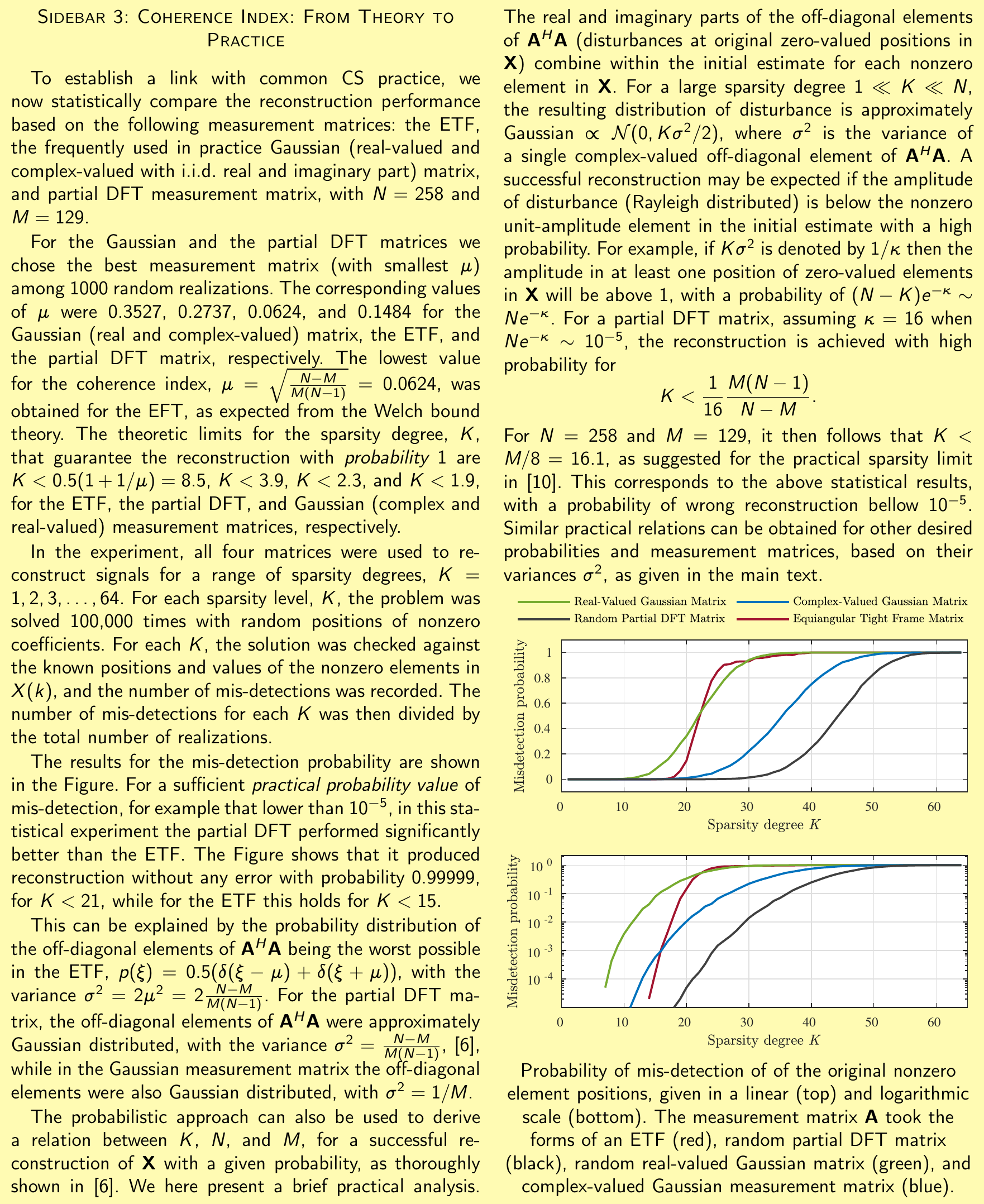}}
\end{figure*}

 Consider now the case with only one nonzero element in $\mathbf{X}$, say at position $k_1=2$, with its initial estimate, $\mathbf{X}_0$ designated by the dark blue bar-plot in Fig. \ref{slika_PPP_1} (top-middle). Therefore, for only one nonzero element in  $\mathbf{X}$, the condition for the correct detection of its position in the estimate $\mathbf{X}_0$ would be that the maximum possible disturbance value, $\mu$, is smaller than the value of the diagonal, which is in this case $\mu<1$.  

Assume next that the signal sparsity index is $K=2$, with nonzero elements  in $\mathbf{X}$ at $X(2)=1$ and $X(7)=1$, as indicated in Fig.  \ref{slika_PPP_1} by the "yellow" and "blue" column in matrix $\mathbf{A}^H \mathbf{A} $.  The set of $M=7$ measurements, $\mathbf{y}=\mathbf{AX}$, is then used to calculate the initial estimate in (\ref{Initi}) as $\mathbf{X}_0=\mathbf{A}^H \mathbf{y}=\mathbf{A}^H \mathbf{A} \mathbf{X}$, as  shown in the stacked bar-plot in Fig.  \ref{slika_PPP_1} (top-right). More specifically, the multiplication of  $\mathbf{A}^H \mathbf{A}$ by $\mathbf{X}$ results in two components in the initial estimate: 
\begin{enumerate}[label=(\roman*)]
\item the component whose values are shown in dark blue bars represents the contribution of  $X(2)$ to the initial estimate $\mathbf{X}_0$, while 
\item  the component designated by yellow bars represents the corresponding contribution of $X(7)$. 
\end{enumerate}

\textit{Remark 4:} Compared to the case with sparsity $K=1$, in the case when the sparsity degree is $K=2$ we can observe two differences: 
\begin{enumerate}[label=(\arabic*)]
\item The disturbances arising from each nonzero element in $\mathbf{X}$ combine, so that the maximum possible disturbance is increased to $2\mu$. 
\item The unit value of the original nonzero element in  $\mathbf{X}$ is also affected by the disturbing values of the other nonzero element in $\mathbf{X}$, with the maximum possible amplitude reduction of this nonzero element of $1-\mu$,  as shown in the stacked bar-plot in Fig. \ref{slika_PPP_1} (top-right).  
\end{enumerate}

 \textit{Remark 5:} From the absolute value of the initial estimate in Fig. \ref{slika_PPP_1} (bottom-right), that is,  $|``\mathrm{blue }\,\, \mathrm{bars}"+``\mathrm{yelow }\,\, \mathrm{bars}"|$, we can conclude that, for $K=2$, the correct nonzero element position in $\mathbf{X}$  will \textit{always} be detected if the original unit amplitude, reduced by the maximum possible disturbance, $\mu$, is greater than the maximum possible disturbance value, $2\mu$, at the original zero-valued positions in  $\mathbf{X}$. In other words, for $K=2$, the reconstruction condition  is given by $1-\mu>2\mu$. If this condition is met, then the position of the nonzero element will always be correctly detected, which is precisely the aim of uniqueness analysis.

Note also that, for rigor, we assumed the worst case scenario for the largest position detection of $X(2)=X(7)=1$, whereby the disturbance from the other nonzero element is the strongest possible. If $|X(2)|>|X(7)|$, this would relax the condition for detection of the $|X(2)|$ position, and vice versa. 

More specifically, the coherence index for this ETF matrix is $\mu=0.2774
$.  In the worst case scenario, the initial estimate values at the nonzero positions $k_1=2$ and $k_2=7$, would be $1-\mu=1-0.2774=0.7226$, which is greater than the largest possible value $2\mu=2 \times 0.2774
=0.5547$ of the initial estimate at the positions where the original vector $\mathbf{X}$ is zero-valued, $k \notin \{2,7\}$. Therefore, the coherence index condition $K<0.5(1+1/\mu)=2.3$ is satisfied for $K=2$. Observe also that the matrix $\mathbf{A}$ defined above cannot be used with $K\ge 3$, since in that case  the maximum possible disturbance of $3\mu$ would be larger than a maximally reduced unit value, at the nonzero element position, in the initial estimate. The amount of this maximum reduction would be $2\mu$, thus resulting in the initial estimate value of $1-2\mu$ and a miss-detection of the nonzero element position in $\mathbf{X}$. 

As an example of the tightness of the coherence index condition, Fig. \ref{coh_text4_1} presents the CS recovery based on a $12 \times 16$ partial DFT measurement matrix, $\mathbf{A}$, with $\mu=0.2455$, for the case of $K=2$. Observe that, unlike in the ETF case, here the disturbing terms are not equal. However, all conclusions regarding the worst case scenario remain valid here and the nonzero element positions in $X(k)$  will be correctly detected based on the presented initial estimate, $X_0(k)$, if $1-\mu>2\mu.$

To provide further intuition, the illustration in Fig. \ref{slika_PPP_1} is next  repeated for a $15 \times 30$ measurement matrix $\mathbf{A}$ and $K=3$, with the results shown in Fig. \ref{slika_PPP_2}. Following the same reasoning as in the previous case, we can conclude from the absolute value of the initial estimate that the detection condition now becomes $1-2\mu>3\mu$. The coherence index for this matrix is therefore $\mu=0.1857$,  with the corresponding coherence index condition $K<0.5(1+1/\mu)=3.2$. Notice that this matrix cannot be used for $K=4$, since $1-3\mu>4\mu$ would not hold, meaning that a disturbance would be misdetected as a nonzero component.

\textit{Remark 6:} Following the above simple and inductive approach, it becomes immediately obvious that, for a general case of a $K$-sparse $\mathbf{X}$, the position of the largest element position in $\mathbf{X}$ will be correctly detected in the initial estimate,  $\mathbf{X}_0$, if
$$1-(K-1)\mu>K\mu.$$
The above bound directly yields the coherence index condition in (\ref{Kkoc}), with the derivation obtained in a natural and practically relevant way.

\textit{Remark 7:} After the position of the first nonzero component in a sparse $\mathbf{X}$ is successfully detected and this component is reconstructed and removed,
the same procedure and relations can be iteratively applied to the remaining ``deflated" signal which now exhibits a reduced 
$(K-1)$-sparsity level, thus guaranteeing a unique solution.

Sidebar 2 provides a simple analytic derivation to support the intuition behind the proposed proof of the condition for unique reconstruction, as illustrated in Fig. \ref{slika_PPP_1}, Fig. \ref{coh_text4_1}, and Fig. \ref{slika_PPP_2}.

The case with small measurement matrices of the EFT type, considered so far, fully supports the proposed derivation of the coherence index and its appropriateness in theory and practice, as it produces a tight bound on the existence and uniqueness of reconstruction. 
The corresponding performance for problems of large dimensions is illustrated in Sidebar 3.

In summary, the coherence index has been derived over several independent  worst case scenarios, to provide a rigorous and easily interpretable bound for practical application scenarios, where the uniqueness condition is typically further relaxed. This this end, in our approach, we considered the case where:
\begin{itemize}
\item \textit{Amplitudes of all nonzero components in $\mathbf{X}$ are equal.} If this is relaxed to the general case of different values of nonzero elements in  $\mathbf{X}$, we can expect a successful, more relaxed and unique reconstructions even if the coherence index condition may be violated.
\item \textit{Only one  element in the initial estimate,  $\mathbf{X}_0$,  was compared to the maximum possible disturbance.} If any of the original nonzero components in the initial estimate at a position of nonzero element in $\mathbf{X}$ is above the maximum disturbance, this would guarantee a successful reconstruction, thus further relaxing the reconstruction  condition.
\item \textit{All disturbances at both the nonzero and zero-valued element positions are assumed to be in phase,} that is, they are summed up with maximum possible magnitudes. This is a very low probability event, again relaxing the practical sparsity limit for the reconstruction.
\item \textit{The distribution of the off-diagonal elements of $\mathbf{A}^H\mathbf{A}$ plays an important role.} While, for several nonzero elements in $\mathbf{X}$, these combine to an approximately Gaussian distributed variable, the resulting disturbance may obey different distribution,  thus yielding different results.           
\end{itemize}

\section{Conclusion}

The coherence index condition for unique sparse signal reconstruction, a prerequisite for the successful compressive sensing paradigm, has been derived using simple signal processing tools and through an intuitive example. Our perspective has first demonstrated that this index provides a tight uniqueness bound for relatively small and moderate dimension of CS problems, followed by a clear interpretation of its conservative nature for large scale problems. This has been achieved for a range of  high probabilities of obtaining the correct result and avoiding mis-detection. It has also been shown that general measurement matrices, like the frequently used partial DFT matrix, are likely to  outperform the results based on the considered ETF based measurement matrix, if the analysis is restricted to pragmatically high probabilities of the correct solution - a practical relaxation of theoretical probability of one for avoiding mis-detection.

\end{document}